\documentclass[final,5p,times,twocolumn]{elsarticle}

\usepackage{amssymb}

\usepackage{amsmath}
\usepackage{amsfonts}

\usepackage{natbib}

\biboptions{square,sort&compress}

\journal{Physics Letters B}

\def\d{\text{d}}
\def\p{\partial}
\def\D#1#2{\frac{\d{#1}}{\d{#2}}}
\def\P#1#2{\frac{\p{#1}}{\p{#2}}}
\def\nno{\nonumber\\}

\def\Si{\text{Si}}
\def\wBD{\omega_{\text{BD}}}

\def\abs#1{\left|#1\right|}

\begin{document}

\begin{frontmatter}



\title{Parametric instability induced scalar gravitational waves from a model pulsating neutron star}


\author[abrd,RAL]{Charles H.-T. Wang}
\ead{c.wang@abdn.ac.uk}

\author[abrd]{Paolo. M. Bonifacio\corref{cor}}
\ead{paolospe76@gmail.com}

\author[RAL,stra]{Robert Bingham}
\ead{bob.bingham@stfc.ac.uk}

\author[IST,RAL]{J. Tito Mendon\c{c}a}
\ead{titomend@ist.utl.pt}


\address[abrd]{SUPA Department of Physics, University of Aberdeen, King's
College, Aberdeen AB24 3UE, UK}
\address[RAL]{STFC Rutherford Appleton Laboratory, Chilton, Didcot, Oxfordshire OX11 0QX, UK}
\address[stra]{SUPA Department of Physics, University of Strathclyde, Glasgow G4 0NG, UK}
\address[IST]{GOLP/Centro de F\'isica de Plasmas, Instituto Superior
T\'ecnico, 1049-001 Lisboa, Portugal}

\begin{abstract}
We identify a new dynamical mechanism for a strong scalar gravitational field effect. To illustrate this mechanism, we investigate the parametric excitation and emission of scalar gravitational waves by a radially pulsating model neutron star.
\end{abstract}


\end{frontmatter}


\section{Introduction}

Within the metric description of gravitation, scalar-tensor (ST)
theory of gravity provides a class of natural extensions to general
relativity (GR), by including a scalar field as part of gravity.
Though a multiplet of scalar fields can be considered, here we shall
focus only on a single real-valued scalar field and denote it by
$\phi$. It modifies gravity in the following way: In the absence of
this field, the motion of matter fields is determined using the
metric $g_{ab}^*$, referred to as the ``Einstein metric'' as per GR.
In the presence of the scalar field $\phi$, the motion of matter
fields are affected by the ``physical metric'' $g_{ab}$ obtained
from the Einstein metric through a conformal mapping $g_{ab} =
A(\phi)^2g_{ab}^*$, where $A(\phi)$ is a theory-dependent ``coupling
function'' \cite{Fierz1956}. Since $A(\phi)$ effectively re-scales
the mass of particles, it is also known as the ``mass function''
\cite{Wagoner1970, Bekenstein1980}. The coupling strength between
matter and scale fields $\phi$ is proportional to the field
derivative $\alpha(\phi):=\p a(\phi)/\p \phi$ of the natural
logarithm $a(\phi):=\ln A(\phi)$ of the coupling function. (See the
field equations below.)

The pioneering scalar-tensor theory by Brans
and Dicke \cite{Brans1961,Dicke1962} is equivalent to the simple
choice of a linear function $a(\phi)= \alpha_0\phi$ where $\alpha_0$
is a constant and is related to the Brans-Dicke parameter $\wBD$ by
$\alpha_0^2=(2\wBD+3)^{-1}$. Thus the larger $\wBD$, the weaker
scalar field coupling. The most stringent test to date of the
Brans-Dicke theory is provided by measuring the frequency shift of
the radio signals to and from the Cassini-Huygens spacecraft,
suggesting that $\wBD > 40000$ \cite{Bertotti2003}. Further tests
could be provided by the measurements of (tensor) gravitational wave
forms from neutron stars and black holes using advanced laser
interferometers \cite{Will1994, Scharre2002}.

The weakness of the Brans-Dicke type scalar coupling can be
explained in terms of the relaxation of a more general nonlinear
function $a(\phi)$ to its local minimum during the cosmological
evolution \cite{Damour1993}. The effect amounts to a local attractor
$\phi_0$ around which $a(\phi) \approx \beta(\phi-\phi_0)^2/2$ with
some constant $\beta > 0$. On the other hand, if $\phi_0$
corresponds to a local maximum, then $\beta < 0$ and ``spontaneous
scalarization'' could occur \cite {Damour1993b, Damour1996}. Indeed,
for  $\beta \lesssim -4$ large deviation of the scalar field from
its cosmological value may be developed inside a neutron star
\cite{Damour1993b, Damour1996}. The implications of this
strong-field effect have attracted considerable recent attention
e.g. \cite{Harada1998, Whinnett2000, Sotani2004}.

The purpose of this paper is to point out a new  strong field effect
for $\beta>0$. The scenario involves dynamical parametric
instability for the scalar field inside a rapidly pulsating strong
source of gravity, such as a young neutron star. In order to gain
some understanding of this new effect, an idealized neutron star
model is analyzed. We show that for some sufficiently large $\beta$
together with certain physically reasonable conditions on the star,
the pulsating model neutron star behaves like an optical cavity in
which  resonant scalar waves are parametrically amplified. The
surface of the star then acts like a partial (anti-phase) reflector
that releases travelling scalar waves analogous to an optical laser.
This work therefore provides an initial estimate of
the effect that can be extended for further investigation with
realistic stars, including the possible energy transfer from a
collapsed star core to stalled shock waves in supernova formations
and other astrophysical problems \cite{Bastrokov2004}.
A further motivation of this work is to seek a possible source of
conformal fluctuations of spacetime as a result of background
scalar gradational waves \cite{Bonifacio2009}.

\section{The scalar gravitational field}

The general action for scalar-tensor gravity including matter can be
expressed as \cite{Wagoner1970, Dicke1962}:
\begin{eqnarray}
S = \frac{c^4}{16\pi G_*} \int\frac{\d^4x}{c} \,g_*^{1/2}R_* +S_\phi
+S_m
\label{act}
\end{eqnarray}
where $a,b,\cdots=0,1,2,3$, $\d^4 x = \d x^0 \d^3 x$ and $x^0 = c
t$, and
\begin{eqnarray}
S_\phi = -\frac{c^4}{4\pi G_*} \int\frac{\d^4x}{c} \,g_*^{1/2}
\left[\frac12\,g^{ab}_*\phi_{,a}\phi_{,b} + V(\phi)\right]
\label{actphi}
\end{eqnarray}
is the action for the scalar field $\phi$ in terms of a potential
function $V(\phi)$ and
$$S_m=S_m[\psi, A^2(\phi)g^*_{ab}]$$
is the
action for matter fields in terms of a coupling function $A(\phi)$.

The Minkowski metric is given by $\eta_{ab}=$ diag$(-1,1,1,1)$.
The effective stress tensor for $\phi$ follows from \eqref{actphi}
as
\begin{eqnarray}
&&T_\phi^{ab} := 2 c\, g_*^{-1/2}\frac{\delta S_\phi}{\delta
g_{ab}^*}= \nno && \frac{c^4}{8\pi G_*} \left[
2g^{ac}_*g^{bd}_*\phi_{,c}\phi_{,d}
-g^{ab}_*(g^{cd}_*\phi_{,c}\phi_{,d} + 2V(\phi)) \right].
\label{Tphi}
\end{eqnarray}

The field equation for $\phi$ follows from varying the total action
$S$ in \eqref{act} as
\begin{eqnarray}
\Box_* \phi - \P{V(\phi)}{\phi} = -\frac{4\pi G_*}{c^4}\, \alpha(\phi)
T_* \label{phieq}
\end{eqnarray}
where $\Box_*$ is the Laplace-Beltrami operator and $T_*$ is the
contracted stress tensor of the matter field with respect to
$g_{ab}^*$.

We shall be concerned with the values of $\phi$ near a local minimum
of $a(\phi)$. Thus up to an additive constant, equivalent to a
re-scaling constant for the metric $g_{ab}$, we have approximately
\begin{eqnarray}
a(\phi) = \frac12\beta\phi^2 \label{A}
\end{eqnarray}
for some constant $\beta>0$. For simplicity we shall consider here
the quadratic potential
\begin{eqnarray}
V(\phi) = \frac12\mu_0^2\phi^2 \label{V}
\end{eqnarray}
that gives rise to an effective mass
$$m_0=\mu_0\hbar/c$$
of the scalar field $\phi$ in vacuum.

Then \eqref{phieq} becomes
\begin{eqnarray}
\Box_* \phi - \mu_0^2\phi = U \phi \label{weq0}
\end{eqnarray}
where
\begin{eqnarray}
U:=-\frac{4\pi G_*\beta}{c^4} T_* . \label{U}
\end{eqnarray}

\section{The model neutron star}

It is interesting to note that \eqref{weq0} is a homogeneous wave
equation for $\phi$. As such, parametric excitation of $\phi$
through a time varying function $U$ could occur. However, generation
of large amplitude scalar wave does not necessarily follow unless
parametric instability takes place. Such a condition would require
rapid variations of a high density gravitational source with a
sufficiently large value for $\beta$. In addition, damping of the
parametric excitation through radiating scalar waves must also be
sufficiently small, otherwise the parametric excitation would be
stabilized.

In this paper, we investigate a particular scenario by
means of a model pulsating neutron star to illustrate how these
conditions can be satisfied through the parametric instability of
the quasi-normal modes of the scalar field inside the star. We
envisage that similar mechanism could be identified in other violent
events in astrophysics and the early universe.

The radiation of scalar gravitational waves by a radially pulsating
star using the Brans-Dicke theory was studied in
\cite{Morganstern1967}. (See also \cite{Wagoner1970}.) The emission
mechanism in that case is essentially the same as that for the
conventional tensor gravitational waves but monopole radiation is
possible. In our present case of quadratic scalar coupling using
\eqref{A}, the dynamical structure is completely different, as we
will see below.

Following \cite{Morganstern1967}, we adopt
$$
T_*=-c^2\rho+3p \label{Tstar}
$$
as the contracted stress tensor for matter inside a spherically
symmetric star model with radius $R$, density $\rho$ and pressure
$p$.

Outside the star we have simply $\rho=0=p$ and hence $T_*=0$.
For this model, we shall further assume $ c^2\rho \gg p $ so that
\eqref{U} can be approximated by
\begin{eqnarray}
U=\frac{4\pi G_* \beta}{c^2}\rho . \label{U1}
\end{eqnarray}

For simplicity, we shall also ignore the curvature of $g_{ab}^*$ by
approximating it with the Minkowski metric $\eta_{ab}$. Therefore
the scalar field equation \eqref{weq0} reduces:
\begin{eqnarray}
\p_0^2\phi - \Delta\phi + \mu_0^2\phi + U\phi = 0 \label{weq}
\end{eqnarray}
where $\Delta$ is the 3-dimensional Laplace operator.

The star is modelled as a solid sphere with a constant equilibrium
density $\rho_0$. Its density fluctuations are described by the wave
equation
\begin{eqnarray}
\p_0^2\rho - (v/c)^2\Delta\rho = 0 \label{req}
\end{eqnarray}
where $v$ is the speed of the density/pressure wave. For simplicity
we shall consider a single mode radial oscillation of the density
subject to zero boundary condition at the surface $r=R$ so that
\begin{eqnarray}
\rho = \rho_0[1 - \epsilon \chi_m(r)\cos(\Omega_m t)] \label{rho}
\end{eqnarray}
for some positive integer $m$ as a mode index, where
\begin{eqnarray}
\Omega_m=\frac{m\pi v}{R} \label{Omega}
\end{eqnarray}
is the oscillation frequency, $\epsilon$ is a dimensionless
amplitude parameter, and
\begin{eqnarray}
\chi_n(r) := \frac{R}{r}\sin(\kappa_n r) \label{fn}
\end{eqnarray}
with the wave number
$$\kappa_n={n\pi}/{R}$$
for $n=1,2,\dots$ The following orthogonality relation holds:
\begin{eqnarray}
\int_{0}^R  \d r \,r^2 \chi_n(r)\chi_m(r)  = \frac{R^3}{2}\delta_{n
m} \label{on}
\end{eqnarray}
for any $m,n=1,2,\dots$

Substituting \eqref{fn} into \eqref{U1} we
have
\begin{eqnarray}
U = U_0[1 - \epsilon \chi_m(r)\cos(\Omega_m t)] \label{VV0}
\end{eqnarray}
where
\begin{eqnarray}
U_0:=\frac{4\pi G_* \beta}{c^2}\rho_0 \ge 0 . \label{U2}
\end{eqnarray}

Then \eqref{weq} becomes
\begin{eqnarray}
\p_0^2\phi - \Delta\phi + \mu^2\phi - \epsilon
U_0\chi_m(r)\cos(\Omega_m t)\phi = 0 \label{vweq}
\end{eqnarray}
where
\begin{eqnarray}
\mu^2 := \mu_0^2 + U_0 . \label{muV}
\end{eqnarray}

The energy density and flux of $\phi$ can then be evaluated from
\eqref{Tphi} using $g_{ab}^*=\eta_{ab}$ to be
\begin{eqnarray}
&&u := T_\phi^{00} = \frac{c^4}{8\pi G_*} \left[
\phi_{,0}\phi_{,0}+\eta^{\alpha\beta}\phi_{,\alpha}\phi_{,\beta} +
2V(\phi) \right] \label{Tphiden}
\\
&&f^\beta:=cT_\phi^{0\beta} = -\frac{c^5}{4\pi G_*}
\phi_{,0}\phi_{,\beta} \label{Tphiflx}
\end{eqnarray}
respectively, where $\alpha,\beta=1,2,3$. The potential takes the
forms
$$V(\phi) = \frac12\mu^2\phi^2$$
and
$$
V(\phi) = \frac12\mu_0^2\phi^2
$$
inside and outside the model star
respectively.

\section{Approximate normal modes of the scalar field inside the
model star}

In the absence of the density oscillation, i.e. $\epsilon=0$, the
model star has a constant density $\rho_0$, inside which $\phi$
satisfies the Klein-Gordon equation with mass parameter $\mu >
\mu_0$ given in \eqref{muV}.

For $\mu/\mu_0 \gg 1$ the star surface
at $r=R$ behaves like a perfect (anti-phase) reflector for outgoing
$\phi$. As such the scalar field inside the star can be approximated
by a standing wave subject to the boundary conditions $\phi(R,t)=0$
as follows:
\begin{eqnarray}
\phi \approx \sum_{n}\phi_n := \sum_{n}\varphi_n(t) \chi_n(r)
\label{anz}
\end{eqnarray}
where $n=1,2\ldots$. Each $\phi_n$ denotes a normal mode with
\begin{eqnarray}
\varphi_n(t) = \Re\, \varphi_{n0} e^{-i\omega_n t} \label{APhik}
\end{eqnarray}
where $ \varphi_{n0}$ is a modal amplitude constant and
\begin{eqnarray}
\frac{\omega_n^2}{c^2} = \kappa_n^2+\mu^2 . \label{wn}
\end{eqnarray}

To evaluate the energy associated with these normal modes, we first
obtain the energy density of $\phi$ inside the model star
\begin{eqnarray}
u =\frac{c^4}{8\pi G_*} [(\phi_{,0})^2+(\phi_{,r})^2 + \mu^2\phi^2]
\label{T00nor}
\end{eqnarray}
by using \eqref{Tphiden} in spherical coordinates.

Using \eqref{fn},
\eqref{on}, \eqref{anz}, \eqref{APhik} and \eqref{wn}, this yields
the following energy for $\phi_n$:
\begin{eqnarray}
E_n = 4 \pi\int_{0}^R  \d r \,r^2 u = \frac{c^2}{2
G_*}R^3\varphi_{n0}^2\omega_n^2 . \label{T00nor1}
\end{eqnarray}

\section{Quasi-normal modes and damping due to transmitted scalar
waves}

For finite $\mu/\mu_0 > 1$, the star surface does allow some scalar
wave to propagate across it.

By taking into account the resulting loss of energy, we can refine
$\phi_n$ in \eqref{anz} to be quasi-normal modes. To this end, we
assume that \eqref{APhik} is valid over a few circles of oscillation
at angular frequency $\omega_n$. Exterior to the star, this yields
the scalar field
\begin{eqnarray}
\phi_n = \Re\, \varphi_{n0}\frac{\kappa_n}{k_n}\frac{R}{r}e^{i(k_n r
- \omega_n t + \theta_n)} \label{phiext}
\end{eqnarray}
exterior to the model star ($r>R$), where $\theta_n$ is a constant
phase and
\begin{eqnarray}
k_n^2 = \frac{\omega_n^2}{c^2} - \mu_0^2 .
\label{wnext}
\end{eqnarray}
Using \eqref{muV}, \eqref{wn} we have
\begin{eqnarray}
\frac{\kappa_n^2}{k_n^2} = \frac{\kappa_n^2}{\kappa_n^2+U_0} .
\label{kapkV}
\end{eqnarray}
The power carried by outgoing waves then follows from
\eqref{Tphiflx} as (at $r \gg R$):
\begin{eqnarray}
P_n = 4\pi r^2 |f^\beta| =
\frac{c^4}{G_*}\frac{\kappa_n^2}{k_n^2}R^2\varphi_{n0}^2\omega_n k_n
\label{pownor}
\end{eqnarray}
This yields the damping factor $d_n$:
\begin{eqnarray}
d_n = \frac{P_n}{E_n} = \frac{2c^2 \kappa_n^2}{R\omega_n k_n}
\label{dr}
\end{eqnarray}
for $\varphi_n$, which now describes a quasi-normal mode satisfying
the damped oscillator equation:
\begin{eqnarray}
\D{^2\varphi_n}{t^2} + d_n \D{\varphi_n}{t} + \omega_n^2 \varphi_n =
0 . \label{meq}
\end{eqnarray}

\section{Parametric excitation of the normal modes}

We now estimate the parametric excitation of the normal modes of the
scalar field in the presence of the density oscillation described by
\eqref{rho} with $\epsilon\neq0$. We shall first neglect the effect
of damping as just discussed and then take this effect into account
later on. In order to gain a simple rough estimate of the excitation
effect we shall further neglect mode coupling through parametric
expiation.

We therefore proceed by applying \eqref{anz} with a single mode for
some $n$ into \eqref{vweq} and then use $\chi_n$ as a test function
to extract the equation for $\varphi_n$, i.e.:
\begin{eqnarray}
\int_{0}^R  \d r \,r^2 \chi_n [(\p_0^2\varphi_n)\chi_n -
\varphi_n\Delta \chi_n + \varphi_n\mu^2 \chi_n\nno - \epsilon
U_0\chi_m(r)\cos(\Omega_m t)\varphi_n \chi_n] = 0 . \label{vweq2}
\end{eqnarray}
Using \eqref{on} this yields
\begin{eqnarray}
\D{^2\varphi_n}{t^2} +\omega_n^2\varphi_n  - \epsilon U_0 c^2
\chi_{nm}\cos(\Omega_m t)\varphi_n = 0 \label{vweq3}
\end{eqnarray}
where
\begin{eqnarray}
&&\chi_{nm}:= \frac2{R^3}\int_{0}^R  \d r \,r^2 \chi_n^2\chi_m =
\nno && \Si(m\pi)-\frac12\Si(2n\pi+m\pi)+\frac12\Si(2n\pi-m\pi) .
\label{rel}
\end{eqnarray}

\section{Modal equation including damping and parametric
excitation}

Incorporating both damping and parametric excitation, we arrive at
the following equation
\begin{eqnarray}
\D{^2\varphi_n}{t^2} + d_n \D{\varphi_n}{t} + \omega_n^2 \varphi_n -
\epsilon c^2 U_0\chi_{nm}\cos(\Omega_m t)\varphi_n= 0
\label{meqn}
\end{eqnarray}
for each quasi-normal mode. To see the stability of these modes we
cast \eqref{meqn} into the following canonical form for damped
Mathieu equation:
\begin{eqnarray}
\D{^2\varphi_n}{\tau^2} + 2\zeta\D{\varphi_n}{\tau} + a \varphi_n -
2q\cos(2\tau)\varphi_n= 0
\end{eqnarray}
by using \eqref{muV}, \eqref{wn}, \eqref{dr}, in terms of the
following dimensionless quantities:
\begin{eqnarray}
\tau&=&\frac{\Omega_m}2 t,\quad \zeta= \frac{2c^2
\kappa_n^2}{R\Omega_m\omega_n k_n}
\nonumber\\
a &=& \frac{4c^2}{\Omega_m^2}(\kappa_n^2+\mu_0^2 + U_0),\quad
q=\frac{2\epsilon c^2 U_0\chi_{nm}}{\Omega_m^2} .
\end{eqnarray}

The stability domain near the principal parametric excitation
frequency with $a \approx 1$ and $q \approx 0$ have been obtained in
\cite{Taylor1969}. For $a=1$, i.e. $\Omega_m = 2\omega_n$,
$\varphi_n$ becomes unstable if the condition
\begin{eqnarray}
\abs{\frac{q}{2\zeta}} &=& \abs{\frac{\epsilon \chi_{nm} R}{4}
\frac{U_0 \sqrt{\kappa_n^2+U_0}}{\kappa_n^2}} \gtrsim 1
\label{instab}
\end{eqnarray}
is satisfied. From \eqref{U2} we see that  the above condition can
be satisfied for sufficiently large $\beta$ and $\epsilon$.

We are grateful to J. Hough (Glasgow), C. L\"ammerzahl (Bremen), J. A. Reid and J. S. Reid (Aberdeen) for helpful discussions, and to the STFC Centre for Fundamental Physics for support. PB acknowledges a Sixth Century Ph.D. Studentship from the University of Aberdeen.

\end{document}